\documentclass[aps,prb,twocolumn,showpacs]{revtex4}
\usepackage{float,graphicx}
\usepackage{dcolumn}
\usepackage{psfig}
 
\newcommand{\bsim}{\mbox{\raisebox{-0.1cm}{$\;
\stackrel{\textstyle>}{\sim}\;$}}}
\newcommand{\lsim}{\mbox{\raisebox{-0.1cm}{$\;
\stackrel{\textstyle<}{\sim}\;$}}}

\begin{document}

\title{Electron-phonon renormalization in small Fermi energy systems}

\author{E. Cappelluti$^{1,2}$,
L. Pietronero$^{2,3}$} 

\affiliation{$^1$``Enrico Fermi'' Center,
v. Panisperna 89a, c/o Compendio del Viminale, 00184 Roma, Italy}

\affiliation{$^2$Dipart. di Fisica, Universit\`{a} di Roma 
``La Sapienza'',  P.le A.  Moro, 2, 00185 Roma,
and INFM UdR Roma1, Italy}

\affiliation{$^3$CNR, Istituto di Acustica ``O.M. Corbino'', v. del Fosso
del Cavaliere 100, 00133 Roma, Italy}

\date{\today}

\begin{abstract}
The puzzling features of recent photoemission data in cuprates
have been object of several analysis in order to identity the
nature of the underlying electron-boson interaction.
In this paper we point out that many basilar assumptions of
the conventional analysis as expected to fail in small Fermi
energy systems when, as the cuprates, the Fermi energy $E_{\rm F}$
is comparable with the boson energy scale.
We discuss in details the novel features appearing in the self-energy
of small Fermi energy systems and the possible implications
on the ARPES data in cuprates.
\end{abstract}

\maketitle

Renewed interest has recently arisen about the role of the electron-phonon
(el-ph)
coupling in cuprates. A powerful tool of investigation
is represented by the
Angle-Resolved Photoemission Spectroscopy (ARPES),
which, in two dimensional systems as the copper oxides, is in principle able
to extract the
electronic self-energy directly from the
momentum and.energy distribution curves.\cite{shen}
The aim of this analysis is
to determine the
microscopic origin and properties of the electron
scattering.\cite{hengsberger,norman,lashell,shi}
For example, in qualitative terms, the report
of a remarkable kink in the
electronic dispersion has been discussed as an evidence of a retarded
electron-boson interaction (most probably of phononic nature), where
the energy at which the kink occurs set the energy scale of the
bosonic spectrum.\cite{lanzara}
In the same framework, the ratio between the slope
of the electronic dispersion at low energy (below the kink) and at high
energy (above the kink) is expected to give a qualitative estimate of
the strength of the electron-boson coupling.\cite{hengsberger,lashell,lanzara}

Some puzzling features question however this conventional el-ph
picture.On one hand, the low energy electronic dispersion shows
a quite weak dependence on the hole doping in contrast with a significant
dependence of the apparent el-ph coupling constant.\cite{zhou}
More interesting, the
{\em high} energy electronic dispersion results to be strongly dependent
on the hole doping.\cite{zhou}
This is particular astonishing because the high energy
part of the electronic dispersion is expected to represent the bare
electronic structure and it should not be dependent on any
electron-phonon properties.\cite{grimvall,engelsberg}
In this situation the attempt to fit this
experimental scenario within a conventional el-ph framework
by means of inversion techniques of the raw data would lead to unrealistic
assumptions for the hypothetical underlying electron-boson
spectrum.\cite{verga,schachinger}

The reliability of a conventional analysis concerning the electron-phonon
properties in cuprates is questioned also on the theoretical ground.
Due to the high degree of the electronic correlation, the charge carriers
in cuprates are characterized by a weakly dispersive
effective band, with a Fermi energy
$E_{\rm F} \simeq 0.3-0.4$ eV.\cite{dagotto}
Similar estimates of $E_{\rm F}$ are obtained by penetration depth
measurements as reported in Uemura's plot.\cite{uemura}
This value should be compared with the highest phonon frequencies
in cuprates $\omega_{\rm ph}^{\rm max} \simeq 80-100$ meV, defining
an adiabatic ratio $\omega_{\rm ph}^{\rm max}/ E_{\rm F} \sim 0.2-0.3$.
In this situation, the adiabatic assumption
($\omega_{\rm ph} \ll E_{\rm F}$),
on which the conventional el-ph analysis relies, breaks down and
novel nonadiabatic interferences between the electronic and lattice degrees
of freedom are expected to affect the normal and superconducting state
phenomenology.\cite{psg}
Among other features, the appearance of nonadiabatic effects
has shown to account in a natural way for the presence of
an finite isotope effect
on the effective electron mass and for
the possibility of high-$T_c$ superconductivity within an el-ph
scenario.\cite{gpsprl,gcp}

Aim of the present paper is to investigate in some detail
how the presence of a small Fermi energy, of the same order of the phonon
frequencies, affects the electron-phonon phenomenology.
As a first step in this direction, we do not consider here the onset
of nonadiabatic vertex diagrams which arise in the nonadiabatic
regime,\cite{psg,gpsprl}
but we retain only the nonadiabatic effects related to the finite
electronic bandwidth. Note however that,
while vertex diagrams play a fundamental role in determining an effective
enhancement of the superconducting pairing,\cite{gpsprl}
finite bandwidth effects alone
have been shown to account in a qualitative way for the anomalous
el-ph effects in different normal state properties (effective mass $m^*$,
Pauli spin susceptibility, ldots)\cite{gcp,cgp}.
We anticipate here, as our main results, that
a conventional (adiabatic) el-ph analysis
needs to be deeply revised in small Fermi energy systems.
In particular, we show that the following fundamental
el-ph properties\cite{grimvall,engelsberg}
are {\em no longer valid}
when $E_{\rm F} \sim \omega_{\rm ph}$:
\begin{itemize}
\item[($i$)]
the el-ph self-energy $\Sigma(\omega)$ does not
renormalize the electronic dispersion for $\omega$ much larger than
the phonon energy scale $\omega_{\rm ph}$;
\item[($ii$)]
impurity scattering affects only the imaginary part of the self-energy
but not the real part, and hence {\em not} the electronic dispersion;
\item[($iii$)]
different channels of electron scattering (phonons, impurities, \ldots)
just sum in the self-energy.
\end{itemize}

Working tools of our analysis will be the Marsiglio-Schossmann-Carbotte
(MSC) equations\cite{msc}
properly generalized in the case of finite bandwidth.
The formal derivation of the MSC iterative procedure in the case of finite
Fermi energy systems follows quite closely the steps outlined
in Refs. \onlinecite{msc,freericks}.
Here we report only the final equations. The procedure can be derived
in full generality for any shape of the density of states (DOS)
and any electron filling. For sake of simplicity,
and in order to disentangle finite bandwidth effects
from the breaking of particle-hole symmetry,
we consider a simple system at half-filling with a constant DOS:
$N(\epsilon)=N(0)$ for $|\epsilon| \le E_{\rm F}$, where
the half-bandwidth represents also the Fermi energy
In the presence of both el-ph interaction and impurity scattering,
MSC equations read thus:
\begin{eqnarray}
\Sigma(i\omega_n)&=&
-2iT \sum_m
\lambda(i\omega_n-i\omega_m)
\eta(\omega_m)
-2i\gamma \eta(\omega_n),
\label{matsu}
\\
\Sigma'(\omega)&=&
2T \sum_m
\lambda'(\omega,\omega_m) \eta(\omega_m)
-\int_{-\infty}^\infty d\Omega
\alpha^2F(\Omega) 
\nonumber\\
&&\times\left[N(\Omega)+f(\Omega-\omega)\right]
\eta'(\omega-\Omega)
- \gamma \eta'(\omega),
\label{real}
\\
\Sigma''(\omega)&=&
-\int_{-\infty}^\infty d\Omega
\alpha^2F(\Omega) \left[N(\Omega)+f(\Omega-\omega)\right]
\nonumber\\
&&\times\eta''(\omega-\Omega)
- \gamma \eta''(\omega),
\label{imaginary}
\end{eqnarray}
where $N(x)$ and $f(x)$ are respectively the Bose
and Fermi distribution functions,
$\alpha^2F(\Omega)$ is the el-ph Eliashberg function,
$\gamma$ the impurity scattering rate. Moreover
$\lambda(z)=
\int_{-\infty}^\infty  d\Omega
\alpha^2F(\Omega)/[\Omega-z]$ ($z$ complex number),
$\lambda'(\omega,\omega_m)= \mbox{Im}\lambda(\omega-i\omega_m)$,
and
\begin{eqnarray*}
\eta(\omega_m)&=&
\arctan\left[\frac{E_{\rm F}}{\omega_m Z(\omega_m)}\right],
\\
\eta'(\omega)
&=&
\frac{1}{2}
\ln\left[\frac{[E_{\rm F}-\omega Z'(\omega)]^2+[\omega Z''(\omega)]^2}
{[E_{\rm F}+\omega Z'(\omega)]^2+[\omega Z''(\omega)]^2}\right],
\\
\eta''(\omega)
&=&
\arctan\left[
\frac{E_{\rm F}-\omega Z'(\omega)}{\omega Z''(\omega)}
\right]
+
\arctan\left[
\frac{E_{\rm F}+\omega Z'(\omega)}{\omega Z''(\omega)}
\right],
\end{eqnarray*}
where $Z(z)=1-\Sigma(z)/z$.
As usual, the self-energy on the real axis is obtained by using the
Matsubara solution [Eq. (\ref{matsu})] as input
in Eqs. (\ref{real})-(\ref{imaginary}).
Note that, in the normal state, this is necessary only as long as
the Fermi energy $E_{\rm F}$ is finite.
Note in addition that also the self-consistency
of Eqs. (\ref{real})-(\ref{imaginary}) is intrinsically related to the
finiteness of $E_{\rm F}$, so that the factors $\eta'(\omega)$,
$\eta''(\omega)$ are functions of the self-energy itself through
the $Z(\omega)$ function. As we are going to see, it is just
the non-linear self-consistent relation of 
Eqs. (\ref{real})-(\ref{imaginary}) which is responsible for the
violation of the conditions ($i$)-($iii$).

\begin{figure}[t]
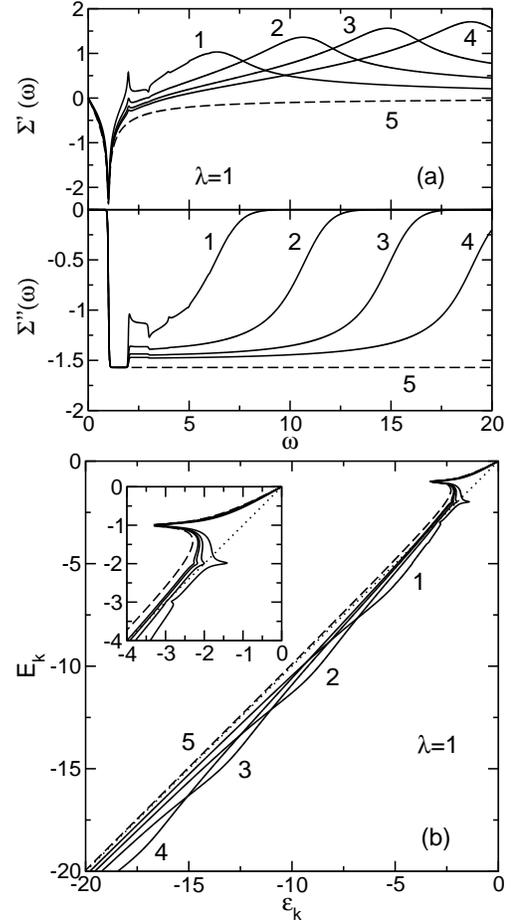

\centerline{\psfig{figure=self-ef.eps,width=6.5cm,clip=}}
\centerline{\psfig{figure=disp-ef.eps,width=6.5cm,clip=}}
\caption{Panel (a): Real and imaginary part of the self-energy for a Einstein
phonon mode with $\lambda=1$ and different Fermi energy. Solid lines
from 1-4: $E_{\rm F}=4,8,12,16\, \omega_0$; Dashed line 5:
$E_{\rm F}=\infty$. All quantities are in units of $\omega_0$.
Panel (b): corresponding renormalized
electron dispersion $E_{\bf k}$ as function of
the bare one $\epsilon_{\bf k}$ (dotted line). In the inset a zoom of the
low energy part.}
\label{f-self-disp-ef}
\end{figure}

In order to underline the small Fermi energy effects on the el-ph
spectral properties, let us consider for the moment a simple Einstein
phonon mode with phonon frequency
$\omega_0$ in the absence of impurities.
In Fig. \ref{f-self-disp-ef}a we show the real and imaginary part
of the self-energy for an Einstein spectrum with
$\lambda= \int_{-\infty}^\infty  d\Omega \alpha^2F(\Omega) /\Omega = 1$
and different Fermi energies.
The dashed lines represent the case of the conventional $E_{\rm F}=\infty$
el-ph: the low energy part of $\Sigma'(\omega)$ is just
$\Sigma'(\omega) \sim \lambda \omega$, which gives the well-known
the renormalization of the electronic dispersion
$E_{\bf k} = \epsilon_{\bf k}/(1+\lambda)$ close to the
Fermi level.\cite{grimvall,engelsberg}
Note that the
real part of the self-energy is always negative implying that
the effective electronic band $E_{\bf k}$
is always less steep than the bare one $\epsilon_{\bf k}$:
$E_{\bf k} \le \epsilon_{\bf k}$ for any energy.
Note also that the magnitude of $\Sigma''(\omega)$ is a monotonously increasing
function with $\omega$ and saturates
for $\omega \ge \omega_{\rm ph}^{\rm max}$.

The presence of a Fermi energy of the same order of the phonon frequencies
gives rise to a number of anomalous features. The most commonly known is the
reduction of the low energy el-ph renormalization 
$E_{\bf k} = \epsilon_{\bf k}/(1+\lambda_{\rm eff})$, where
$\lambda_{\rm eff} = - \lim_{\omega \rightarrow 0} \Sigma'(\omega)/\omega$ is
shown, already by a simple Matsubara analysis, to be less than $\lambda$
due to finite bandwidth effects. More interesting, we note two
{\em qualitatively} new features which appear for $\omega_0/ E_{\rm F} \neq 0$.
The first one is that $\Sigma''(\omega)$ is no longer a monotonous function
of $\omega$, but when $\omega$ becomes roughly 
$\omega \bsim E_{\rm F}$
the imaginary part of the self-energy starts to decrease
and it goes quite rapidly to zero. This is easily
understandable in small Fermi energy systems if one considers that
for $\omega \gg E_{\rm F}$
there are no electronic states into which an electron
with energy $\omega$ could decay within an energy window $\sim \omega_0$.
Another interesting feature is the large positive hump of the real part
of the self-energy occurs by the Kramers-Kronig relations
in correspondence of the drop of the imaginary part and it scales
with $E_{\rm F}$.
In particular we note that, in contrast with the case $E_{\rm F}=\infty$,
for finite $E_{\rm F}$ the real part
of the self-energy $\Sigma'(\omega)$ becomes positive in a large range of
energy for $\omega \bsim 2 \omega_0$. The positiveness of $\Sigma'(\omega)$
has important consequences on the renormalized electronic dispersion
obtained by $E_{\bf k} - \epsilon_{\bf k} - \Sigma'(E_{\bf k})=0$
which corresponds in ARPES measurements to the dispersion inferred by
the momentum distribution curves (MDC). As shown in 
Fig. \ref{f-self-disp-ef}b the positive part of $\Sigma'(\omega)$
implies an ``anti-rinormalization'' of the electron band, namely
$E_{\bf k} > \epsilon_{\bf k}$. This new feature extends up to an energy
scale which does not depend on $\omega_0$ but only on $E_{\rm F}$, while
its magnitude depends on el-ph parameters as $\lambda$ or $E_{\rm F}$
itself. In such a situation the high energy part
$E_{\bf k} > \omega_{\rm ph}$ of the experimental
electronic dispersion\cite{lanzara,zhou}
does not represent anymore the bare band
$\epsilon_{\bf k}$ but it expected to show a {\em steeper} behavior
than $\epsilon_{\bf k}$. As a last observation, note the kinks/jumps in the
real and imaginary parts of the self-energy occurring at the multiples
of $\omega_0$. These anomalies were predicted already in
Ref. \onlinecite{engelsberg}

As we have just shown, in small Fermi energy system particular care is needed
in order to disentangle el-ph properties from the knowledge of
the renormalized electronic dispersion as one could get from ARPES.
This issue is hardened by the fact that other actors can play
an important role on the renormalization of the electron band.
Most important is the presence of disorder and impurities.
In conventional metals where $E_{\rm F}$ is much larger than any other
energy scale impurity scattering can be considered in good approximation as
static: it provides a finite quasi-particle lifetime in the imaginary
part of the self-energy, but it does not affect the real part
of $\Sigma$, and consequently the electronic dispersion which can
be thought to be determined only by the retarded (boson mediated) scattering.
Things are different in small Fermi energy systems.

\begin{figure}
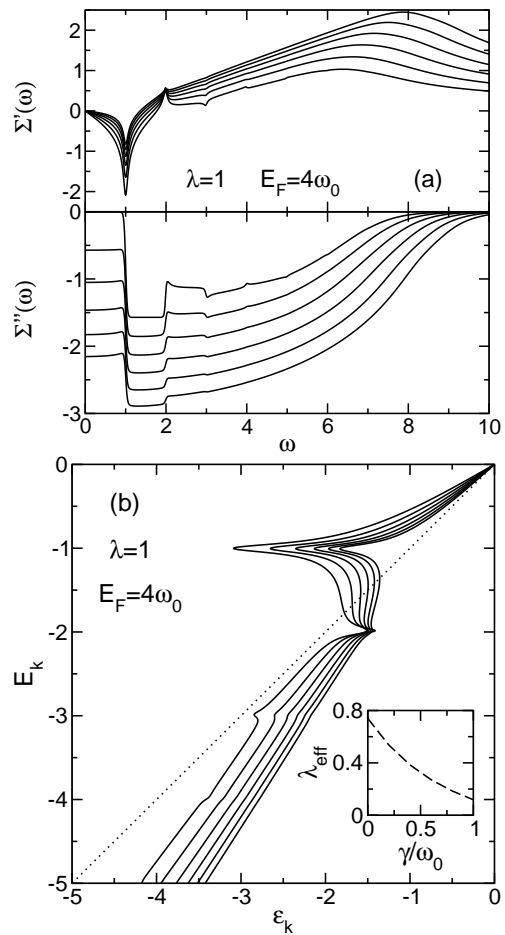

\centerline{\psfig{figure=self-g.eps,width=6.5cm,clip=}}
\centerline{\psfig{figure=disp-g.eps,width=6.5cm,clip=}}
\caption{Panel (a): Real and imaginary part of $\Sigma$ for a Einstein
phonon mode with $\lambda=1$ and $E_{\rm F}=4 \omega_0$
in the presence of impurity scattering.
Solid lines corresponds to (upper panel: from bottom to the top;
lower panel: from top to the bottom):
$\gamma/\omega_0=0, 0.2,0.4, \ldots, 1.0$ where $\gamma$ is the impurity
scattering rate. Energy quantities are expressed in units of $\omega_0$.
Panel (b): renormalized
electron dispersion corresponding (from left to the right) to panel (a).
Inset: dependence of the effective parameter $\lambda_{\rm eff}$ on the
impurity scattering rate.}
\label{f-self-disp-g}
\end{figure}

In Fig. \ref{f-self-disp-g}a we plot the real and imaginary part of the
self-energy for an el-ph Einstein model with $E_{\rm F}=4 \omega_0$
and $\lambda=1$ in the presence of impurity scattering.
In the contrast with the $E_{\rm F}=\infty$ case we see that
the impurity scattering has important effects on the real part of $\Sigma$.
On one hand it smooths the el-ph resonance at $\omega=\omega_0$ as well
as the additional ones at $\omega=n\omega_0$. On the other hand
it significantly enhances the positive part of $\Sigma'(\omega)$ for
$\omega \gg \omega_0$.
The difficulty to extract in a correct way information about
the el-ph spectrum $\alpha^2 F(\Omega)$ from the (ARPES-like)
$E_{\bf k}$ vs. $\epsilon_{\bf k}$ is thus even higher than in the absence
of impurities, as we show in
Fig. \ref{f-self-disp-g}b.
In particular one should take into account that
neither the magnitude of the kink at $\omega=\omega_0$
neither its broadness are directly related anymore 
to the properties of the el-ph spectrum
$\alpha^2 F(\Omega)$ like its total strength
$\sim \lambda$ and its frequency shape.
To be more specific, even the low and high energy parts
of $E_{\bf k}$ are affected in a remarkable way by the
impurity scattering. In the inset we plot
the dependence of the ``effective''
$\lambda_{\rm eff} = - \lim_{\omega \rightarrow 0} \Sigma'(\omega)/\omega$ 
as function of the impurity scattering rate. We note thus that
if the slope of the low energy part were used to extract the el-ph coupling,
we would get a 
strong underestimation of $\lambda$
In addition, the discrepancy between $E_{\bf k}$ and the bare
dispersion $\epsilon_{\bf k}$ at high energy is even larger
in the presence of impurities so that a steeper bare electron dispersion
than the real one would be predicted if the high energy part of
$E_{\bf k}$ to estimate it.
For instance, for $E_{\rm F} = 4 \omega_0$,
$\lambda=1$ and $\gamma/\omega_0=1$ we would get
$E_{\bf k}^{\rm high} \sim 1.6 \,\epsilon_{\bf k}$.

\begin{figure}
\centerline{\psfig{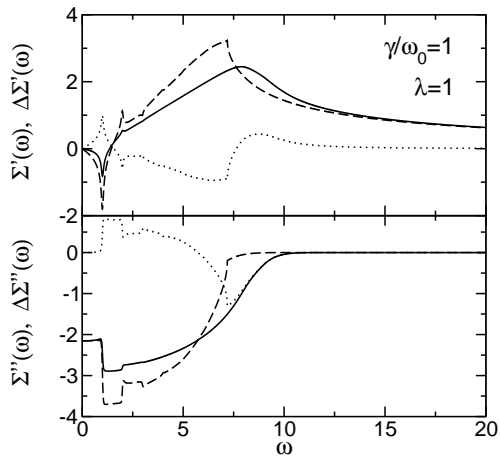}}
\caption{
Real and imaginary part of the self-energy $\Sigma_{\rm el-ph + imp}$
an el-ph + impurities system with $\lambda=1$, $\gamma=\omega_0$
and $E_{\rm F}=4 \omega_0$ (solid line), as compared
with the self-energy obtained
by summing independent contributions
$\Sigma_{\rm el-ph}*\Sigma_{\rm imp}$ (dashed line).
The dotted line represents
the difference $\Delta\Sigma=\Sigma_{\rm el-ph + imp} -
\Sigma_{\rm el-ph}-\Sigma_{\rm imp}$
which is related to the non-linear feedback
arising in small Fermi energy systems.}
\label{f-dself}
\end{figure}

As a last point in this paper, we would like to stress the non-linear and
self-consistent nature of the normal-state
MSC equations (\ref{real})-(\ref{imaginary})
in the presence of a finite Fermi energy. In fact, when
$E_{\rm F} \rightarrow \infty$, $\eta'(\omega)= 0$,
$\eta''(\omega)= \pi \mbox{sgn}(\omega)$, 
the right-hand side of
Eqs. (\ref{real})-(\ref{imaginary})
do not depend anymore on the function $Z(\omega)$,
and the different channels of interaction (el-ph, impurities, spin
fluctuations etc\ldots) just sum up linearly, so that
$\Sigma(\omega)=\Sigma_{\rm el-ph}(\omega)+\Sigma_{\rm imp}(\omega)
+\Sigma_{\rm spin}(\omega) + \ldots$, where each term can be evaluated in the
absence of the other interaction channels.
Due to the self-consistent feedback of $\eta'$ and $\eta''$, this is no more
true in small Fermi energy systems, and a simplistic analysis where
the self-energy was fitted by a sum of independent different contributions
is expected to fail. To illustrate explicitly this point we show in
Fig. \ref{f-dself} the total self-energy $\Sigma_{\rm el-ph + imp}$
for an el-ph + impurities system with $\lambda=1$, $\gamma=\omega_0$
and $E_{\rm F}=4 \omega_0$, 
in comparison with the self-energy obtained
as sum of independent contributions
$\Sigma_{\rm el-ph}+\Sigma_{\rm imp} \equiv
\Sigma(\lambda=1,\gamma/\omega_0=0)+\Sigma(\lambda=0,\gamma/\omega_0=1)$.
As we see, the discrepancies between the fully self-consistent self-energy
and its approximation as sum of two independent contributions can be
quite important in the whole range of energy $\omega$.
In particular we note in the imaginary part that the
jump at $\omega=\omega_0$ due to the onset of el-ph decay processes can be
much smaller than what expected by the simple sum of independent el-ph
and impurity contributions. This observation thus questions the
possibility to estimate the el-ph coupling by
difference between $\Sigma_{\rm el-ph}(\omega) > \omega_{\rm ph}^{\rm max})$
and $\Sigma_{\rm el-ph}(\omega=0)$ if the other contributions could be
disentangled,
in the contrast
with the $E_{\rm F}=\infty$ case where this difference is simply
related to $\lambda \langle \omega \rangle$.
We also note however that, due to the vanishing of the el-ph
processes, the limit $\omega \rightarrow 0$ of the self-energy is not
affected by the interplay between the different channels of interaction,
but it is mainly determined by the only impurity scattering.
It can be still thus used in a safe way to estimate $\gamma$.

In conclusion, motivated by the puzzling features of recent
ARPES measurements in the cuprates,
in this paper we have revised the el-ph properties in
small Fermi energy systems with $E_{\rm F}$ of the same order of the
phonon frequencies, and the interplay with other kinds of scattering,
as impurity-like.
We find that many basilar assumptions which are valid in conventional
metals where $E_{\rm F} \gg \omega_{\rm ph}, \gamma$ need to be strongly
reconsidered. In more details we have shown that in small Fermi energy:
\begin{itemize}
\item
a positive part of the self-energy is predicted for $\omega
\bsim \omega_{\rm ph}$. This implies
an ``anti-rinormalization''
of the electronic band for $\omega_{\rm ph} \lsim \omega
\lsim E_{\rm F}$. ARPES data are expected to measure thus in this regime
an electronic dispersion steeper than the bare one;
\item
impurity scattering can significantly affect the real part of the
self-energy and hence the electronic dispersion.
In particular the effective renormalization of low energy part
of the electronic dispersion is reduced by the impurity scattering
which can also ``wash out'' the el-ph kink
at $\omega \simeq \omega_{\rm ph}$;
\item
different scattering channels are not additive as in the case
of conventional systems. Fitting data by using a sum  of independent
self-energy contribution is expected to fail.
\end{itemize}
Taking into account all these effects in a compelling way would make more
complex but also more interesting the analysis of ARPES data in cuprates
and it could
contribute to understand many puzzling features which are still open issues
in these compounds. We would like to finally stress
that the present analysis is not restricted to the el-ph case, but it applies
equally well for any retarded interaction independently of the specific origin
of the bosonic mediator.

This work was partially supported by PRA-UMBRA project
of INFM and FIRB RBAU017S8R of MIUR.

\end{document}